\documentstyle[12pt]{article}
\newcommand{\lsim}{\raisebox{0.3mm}{\em $\, <$}
\hspace{-3.3mm} \raisebox{-1.8mm}{\em $\sim \,$}}
\newcommand{\gsim}{\raisebox{0.3mm}{\em $\, >$}
\hspace{-3.3mm} \raisebox{-1.8mm}{\em $\sim \,$}}
\begin{document}
\rightline{TMUP-HEL-9604}
\rightline{February 1996}
\baselineskip=19pt
\vskip 0.7in
\begin{center}
{\large{\bf THREE FLAVOR NEUTRINO OSCILLATION CONSTRAINTS FROM}}
{\large{\bf ACCELERATOR, REACTOR, ATMOSPHERIC AND SOLAR NEUTRINO EXPERIMENTS}}
\end{center}
\vskip 0.4in
\begin{center}
Osamu Yasuda\footnote{Email: yasuda@phys.metro-u.ac.jp}
and Hisakazu Minakata\footnote{Email: minakata@phys.metro-u.ac.jp}
\vskip 0.2in
{\it Department of Physics, Tokyo Metropolitan University}

{\it 1-1 Minami-Osawa Hachioji, Tokyo 192-03, Japan}
\end{center}

\vskip .7in
\centerline{ {\bf Abstract} }

We discuss constraints on three flavor neutrino mixings from the
accelerator and reactor experiments, the Kamiokande multi-GeV data,
and the solar neutrino observations. The LSND result is excluded at
90\%CL by the constraints imposed by all the data of reactor and
accelerator experiments and the Kamiokande multi-GeV data if the mass
scale required for the solution to the solar neutrino problem is
hierarchically small. The region of a set of the effective two-flavor
mixing parameters ($\Delta m^2$, $\sin^22\theta$) is given for the
channel $\nu_\mu\rightarrow\nu_e$ which is allowed at 90\%CL by the
multi-GeV Kamiokande data alone.
\newpage

Recently the LSND collaboration has claimed that they have found candidate
events for ${\bar \nu_\mu}\rightarrow{\bar \nu_e}$ oscillation \cite{lsnd}
(See also \cite{hill}).  If their result turns out to be correct,
it gives us an important information for masses and mixing angles
of neutrinos.  In this paper a possibility is explored that all the
experimental data, solar, atmospheric, reactor and accelerator data
including LSND can be explained within a framework of three flavor 
neutrino mixing. It turns out that the LSND result is excluded at 
90\%CL by the constraint imposed by all the data of reactor and 
accelerator experiments and the Kamiokande multi-GeV data. The 
statement is true even without invoking actual solar neutrino data 
as far as the mass scale required for the solution is hierarchically 
small.  

Among the atmospheric neutrino data \cite{kamioka1} \cite{kamioka2} 
\cite{imb} \cite{nusex} \cite{frejus} \cite{soudan2} we discuss only 
the Kamiokande multi-GeV data \cite{kamioka2} throughout this paper.
The reason for this restriction is three-fold; (1) It is the only 
experiment that gives us a nontrivial zenith-angle dependence which 
leads to the upper and lower bounds for the mass-squared difference 
of neutrinos. (2) The result of Monte-Carlo simulation for neutrino 
energy spectrum is published only for the the Kamiokande 
multi-GeV data. (3) It seems difficult to reconcile it with NUSEX 
\cite{nusex} and Frejus \cite{frejus} data. 

The Dirac equation for three flavors of neutrinos in vacuum is given by
\begin{eqnarray}
i{d{ } \over dx} \Psi(x) = U{\rm diag}(E_1,E_2,E_3) U^{-1}\Psi(x),
\end{eqnarray}
which is easily solved:
\begin{eqnarray}
\Psi_\alpha(x)=\sum_{j=1}^3 U_{\alpha j}U^\ast_{\beta j}e^{iE_j x}
\Psi_\beta(x),
\end{eqnarray}
where $E_j\equiv\sqrt{p^2+m_j^2}~(j=1,2,3)$ is the energy of neutrinos
in the mass basis, $\Psi_\alpha(x)=(\nu_e(x),~\nu_\mu(x),~\nu_\tau(x))^T$
~($\alpha$=$e,\mu,\tau$) is the wave function of neutrinos in the
flavor basis, and 
\begin{eqnarray}
U&\equiv&\left(
\begin{array}{ccc}
U_{e1} & U_{e2} &  U_{e3}\\
U_{\mu 1} & U_{\mu 2} & U_{\mu 3} \\
U_{\tau 1} & U_{\tau 2} & U_{\tau 3}
\end{array}\right)\nonumber\\
\displaystyle&\equiv&\left(
\begin{array}{ccc}
c_{12}c_{13} & s_{12}c_{13} &   s_{13}\\
-s_{12}c_{23}-c_{12}s_{23}s_{13} & c_{12}c_{23}-s_{12}s_{23}s_{13}
& s_{23}c_{13}\\
s_{12}s_{23}-c_{12}c_{23}s_{13} & -c_{12}s_{23}-s_{12}c_{23}s_{13}
& c_{23}c_{13}
\end{array}
\right)
\end{eqnarray}
with $c_{ij}\equiv\cos\theta_{ij},~s_{ij}\equiv\sin\theta_{ij}$
is the orthogonal mixing matrix of neutrinos.
We will not discuss the CP violating phase of the mixing matrix
here for simplicity.  The probability of $\nu_\alpha\rightarrow\nu_\beta$
transition is given by
\begin{eqnarray}
&{ }&P(\nu_\alpha\rightarrow\nu_\beta;E,L)\nonumber\\
&{ }&\equiv\left\{ \begin{array}{lr}
\displaystyle
1-4\sum_{i<j}\sin^2\left({\Delta E_{ij}L \over 2}\right)U_{\alpha i}^2
U_{\alpha j}^2\quad&{\rm for}~\alpha=\beta\\
\displaystyle-4\sum_{i<j}\sin^2\left({\Delta E_{ij}L \over 2}\right)
U_{\alpha i} U_{\alpha j} U_{\beta i} U_{\beta j}
\quad&{\rm for}~\alpha\ne\beta,
\end{array} \right.
\end{eqnarray}
where $\Delta E_{ij}\equiv E_i-E_j\simeq (m_i^2-m_j^2)/2E\equiv
\Delta m_{ij}^2/2E$ is the difference of the energy of two mass
eigenstates.

The number of neutrinos $\nu_\beta$ ($\beta=e,\mu,\tau$) is measured
in terms of charged leptons $\ell_\beta$ which come out from a scattering
$\nu_\beta X\rightarrow \ell_\beta X'$.
In appearance experiments of $\nu_\alpha\rightarrow \nu_\beta$
the expected number of the charged leptons $\ell_\beta$
is given by
\begin{eqnarray}
\displaystyle
&{ }&N(\nu_\alpha\rightarrow\ell_\beta;L)\nonumber\\
&=&n_T\int_0^\infty dE
\int^{q_{\rm max}}_0 dq
~\epsilon (q)
F_\alpha (E) { d\sigma_\beta (E,q) \over dq } 
P(\nu_\alpha\rightarrow\nu_\beta; E,L),
\end{eqnarray}
whereas in disappearance experiments we measure attenuation of beam 
\begin{eqnarray}
\displaystyle
&{ }&N_{\alpha\alpha} (L) - N(\nu_\alpha\rightarrow\ell_\alpha;L)\nonumber\\
&=&n_T \int_0^\infty dE
\int^{q_{\rm max}}_0 dq
~\epsilon (q)
F_\alpha (E) { d\sigma_\alpha (E,q) \over dq } 
\left( 1-P(\nu_\alpha\rightarrow\nu_\alpha; E,L)\right),
\end{eqnarray}
where
\begin{eqnarray}
N_{\alpha\beta}(L)\equiv n_T\int_0^\infty dE
\int_0^{q_{\rm max}} dq~\epsilon(q)F_\alpha (E)
{d\sigma_\beta(E,q) \over dq}.
\end{eqnarray}
$F_\alpha (E)$ is the flux of neutrino $\nu_\alpha$ with energy
$E$, $n_T$ is the number of target nucleons, $\epsilon(q)$ is 
the detection efficiency function for charged leptons $\ell_\beta$ 
of energy $q$, $d\sigma_\beta(E,q)/dq$ is the differential cross 
section of the interaction $\nu_\beta X\rightarrow\ell_\beta X'$,
$P(\nu_\alpha\rightarrow\nu_\beta; E)
\equiv|\langle\nu_\alpha (0)|\nu_\beta(L)\rangle|^2$ is the
probability of $\nu_\alpha\rightarrow\nu_\beta$ transitions with
energy $E$ after traveling a distance $L$.  The results of experiments
are usually expressed in terms of a set
of the oscillation parameters ($\Delta m^2,\sin^22\theta$)
in the two-flavor analysis.  The
probability in the two-flavor mixing is given by
\begin{eqnarray}
&{ }&P(\nu_\alpha\rightarrow\nu_\beta;E,L)\nonumber\\
&{ }&\equiv\left\{ \begin{array}{lr}
\displaystyle
1-\sin^22\theta_\alpha\sin^2\left({\Delta m^2L \over 4E}\right)
\quad&{\rm for}~\alpha=\beta\\
\displaystyle
\sin^22\theta_{\alpha\beta}\sin^2\left({\Delta m^2L \over 4E}\right)
\quad&{\rm for}~\alpha\ne\beta.
\end{array} \right.
\end{eqnarray}
Introducing the notation
\begin{eqnarray}
&{ }&\left\langle\sin^2\left({\Delta m^2L \over 4E}\right)
\right\rangle_{\alpha\rightarrow\beta}
\nonumber\\
&\equiv&{1 \over
N_{\alpha\beta}(L)}n_T \int_0^\infty dE
\int_0^{q_{\rm max}} dq~\epsilon(q)F_\alpha (E)
{d\sigma_\beta(E,q) \over dq}
\sin^2\left({\Delta m^2L \over 4E}\right),
\end{eqnarray}
we have the expected number of charged leptons in the two-flavor scenario
\begin{eqnarray}
1-{N^{(2)}(\nu_\alpha\rightarrow\ell_\alpha;L)
\over N_{\alpha\alpha}(L)}
=\sin^22\theta_\alpha(\Delta m^2)\left\langle
\sin^2\left({\Delta m^2L \over 4E}\right)
\right\rangle_{\alpha\rightarrow\alpha} \nonumber\\
\hfill{\rm (disappearance~experiments)}\nonumber\\
{N^{(2)}(\nu_\alpha\rightarrow\ell_\beta;L) \over N_{\alpha\beta}(L)}
=\sin^22\theta_{\alpha\beta}(\Delta m^2)\left\langle\sin^2
\left({\Delta m^2L \over 4E}\right)
\right\rangle_{\alpha\rightarrow\beta} \nonumber\\
\hfill{\rm (appearance~experiments)},
\end{eqnarray}
where we have denoted the $\Delta m^2$ dependence of the mixing angle
$\theta(\Delta m^2)$ explicitly to indicate that $\theta(\Delta m^2)$
is a function of $\Delta m^2$ in the two-flavor analysis.

The boundary of the excluded region (for negative results) in the 
($\Delta m^2,\sin^22\theta$) plot is determined by 
\begin{eqnarray}
\epsilon = {N^{(2)}(\nu_\alpha\rightarrow\ell_\beta;L)
\over N_{\alpha\beta}(L)}
= \sin^22\theta_{\alpha\beta}(\Delta m^2)
\left\langle
\sin^2\left({\Delta m^2L \over 4E}\right)
\right\rangle_{\alpha\rightarrow\beta},
\label{eqn:2flavor1}
\end{eqnarray}
for appearance experiments where the charged leptons are detected 
at one point at a distance $L$, or
\begin{eqnarray}
\hspace{-4.3mm}\epsilon&=&{N^{(2)}(\nu_\alpha\rightarrow\ell_\alpha;L_1)
\over N_{\alpha\alpha}(L_1)}
-{N^{(2)}(\nu_\alpha\rightarrow\ell_\alpha;L_2) \over N_{\alpha\alpha}(L_2)}
\nonumber\\
&=&\sin^22\theta_\alpha(\Delta m^2)
\left[\left\langle
\sin^2\left({\Delta m^2L_2\over 4E}\right)
\right\rangle_{\alpha\rightarrow\alpha}
\hspace{-2.3mm}-\left\langle
\sin^2\left({\Delta m^2L_1\over 4E}\right)
\right\rangle_{\alpha\rightarrow\alpha}\right]
\label{eqn:2flavor2}
\end{eqnarray}
for disappearance experiments 
where the charged leptons are detected at two points
at distances $L_1$ and $L_2$ ($L_1<L_2$).
In (\ref{eqn:2flavor1}) and (\ref{eqn:2flavor2}) 
$\epsilon$ denotes the largest fraction of the appearance events
allowed by a given confidence level, i.e., 
$N^{(2)}(\nu_\alpha\rightarrow\ell_\beta;L)/N_{\alpha\beta}(L)< \epsilon$
for appearance experiments, and 
the largest fraction of beam attenuation, 
$|N^{(2)}(\nu_\alpha\rightarrow\ell_\alpha;L_1)/N_{\alpha\alpha}(L_1)
-N^{(2)}(\nu_\alpha\rightarrow\ell_\alpha;L_2)/N_{\alpha\alpha}(L_2)|
< \epsilon$ for disappearance experiments.

From (\ref{eqn:2flavor1}) and (\ref{eqn:2flavor2}), we can 
read off the value of $\left\langle
\sin^2\left(\Delta m^2L/4E\right)\right\rangle$
for {\it arbitrary} $\Delta m^2$ from the figure
of the two-flavor mixing parameters ($\Delta m^2,\sin^22\theta$)
given in each experimental paper as long as
$\sin^22\theta(\Delta m^2)\le 1$\footnote{In case of the CP violating
phase, which will not be discussed in the present paper, this is not the case.
To discuss the CP violating effect, one needs the information
of $\left\langle\sin\left({\Delta m^2L \over 2E}\right)
\right\rangle$, which cannot be obtained from the
information in published papers only.}:
\begin{eqnarray}
\left\langle
\sin^2\left({\Delta m^2L_2\over 4E}\right)
\right\rangle_{\alpha\rightarrow\alpha}
-\left\langle
\sin^2\left({\Delta m^2L_1\over 4E}\right)
\right\rangle_{\alpha\rightarrow\alpha}
= {\epsilon \over \sin^22\theta_\alpha(\Delta m^2)}\nonumber\\
\hfill{\rm (disappearance~experiments)}\nonumber\\
\left\langle
\sin^2\left({\Delta m^2L \over 4E}\right)
\right\rangle_{\alpha\rightarrow\beta}
= {\epsilon \over \sin^22\theta_{\alpha\beta}(\Delta m^2)}\nonumber\\
\hfill{\rm (appearance~experiments)}.
\label{eqn:average}
\end{eqnarray}
In (\ref{eqn:average}) and in the following,
$\sin^22\theta_\alpha(\Delta m^2)$ or
$\sin^22\theta_{\alpha\beta}(\Delta m^2)$ stands for the value
of $\sin^22\theta(\Delta m^2)$ on the boundary of the
allowed region in the ($\Delta m^2,\sin^22\theta$) plot in the two-flavor
analysis.

Similarly we can express the number
of the expected charged leptons in the three flavor mixing:
\begin{eqnarray}
1-{N^{(3)}(\nu_\alpha\rightarrow\ell_\alpha;L) \over N_{\alpha\alpha}(L)}
=4\sum_{i<j} U_{\alpha i}^2 U_{\alpha j}^2
\left\langle\sin^2\left({\Delta m_{ij}^2L \over 4E}\right)
\right\rangle_{\alpha\rightarrow\alpha}
\qquad\quad{\rm for}~\nu_\alpha\rightarrow\nu_\alpha\nonumber\\
{N^{(3)}(\nu_\alpha\rightarrow\ell_\beta;L) \over N_{\alpha\beta}(L)}
=-4\sum_{i<j}
U_{\alpha i} U_{\alpha j} U_{\beta i} U_{\beta j}
\left\langle\sin^2\left({\Delta m_{ij}^2L \over 4E}\right)
\right\rangle_{\alpha\rightarrow\beta}
~{\rm for}~\nu_\alpha\rightarrow\nu_\beta.
\label{eqn:3flavor}
\end{eqnarray}
From (\ref{eqn:average}) we observe that
the quantity $\left\langle\sin^2\left(\Delta m^2L/4E\right)
\right\rangle/\epsilon$ is equal to $\sin^22\theta(\Delta m^2)$ which
can be read off from the published literatures, and we can express the
conditions for the three flavor mixing parameters in case of negative
results:
\begin{eqnarray}
\epsilon > 
{N^{(3)}(\nu_\alpha\rightarrow\ell_\alpha;L_1)
\over N_{\alpha\alpha}(L_1)}
-{N^{(3)}(\nu_\alpha\rightarrow\ell_\alpha;L_2) \over N_{\alpha\alpha}(L_2)} =
\displaystyle\sum_{i<j} {\epsilon \over \sin^22\theta_\alpha(\Delta m_{ij}^2)}
U_{\alpha i}^2 U_{\alpha j}^2\nonumber\\
\hfill{\rm (disappearance~experiments)}\nonumber\\
\epsilon > {N^{(3)}(\nu_\alpha\rightarrow\ell_\beta;L)
\over N_{\alpha\beta}(L)}=
\displaystyle-4\sum_{i<j} {\epsilon \over \sin^22\theta_{\alpha\beta}
(\Delta m_{ij}^2)}
U_{\alpha i} U_{\alpha j} U_{\beta i} U_{\beta j}\nonumber\\
\hfill{\rm (appearance~experiments)}.
\end{eqnarray}
Notice that the left-hand side of
(\ref{eqn:average}) is defined independent of the number of
flavors of neutrinos.

Throughout this paper we assume that a single mass scale is involved 
in the solution of the solar neutrino problem, which is 
hierarchically small compared to others. Namely, we assume that 
\begin{eqnarray}
\Delta m_{21}\ll \Delta m_{32}^2<\Delta m_{31}^2,
\label{eqn:hierarchy}
\end{eqnarray}
where we have assumed $m_1^2<m_2^2<m_3^2$ without loss of generality.
In the case where $\Delta m_{32}\ll \Delta m_{21}^2<\Delta m_{31}^2$,
we can show that we obtain the same conclusions, although we will not
give the calculation here.

The hierarchy (\ref{eqn:hierarchy}) is satisfied in the two-flavor 
mixing solution to the solar neutrino 
problem \cite{solar} \cite{msw} which requires,
\begin{eqnarray}
&{ }&(\Delta m^2,\sin^22\theta)\nonumber\\
&\simeq&(\Delta m_{21}^2,\sin^22\theta_{12})_\odot\nonumber\\
&\equiv&\left\{ \begin{array}{lr}
({\cal O}(10^{-11}{\rm eV}^2),{\cal O}(1)),&  
({\rm vacuum~solution})\\
({\cal O}(10^{-5}{\rm eV}^2),{\cal O}(10^{-2})),&
({\rm small~angle~MSW~solution})\\
({\cal O}(10^{-5}{\rm eV}^2),{\cal O}(1))&
({\rm large~angle~MSW~solution}).
	     \end{array} \right.
\label{eqn:solar}
\end{eqnarray}
These mass scales are much smaller than those which appear in the 
atmospheric neutrino observations \cite{kamioka2}, or in the LSND 
experiment \cite{lsnd}.  In fact the only condition on
$(\Delta m_{21}^2,\sin^22\theta_{12})$
which is essential to the discussions below is (\ref{eqn:hierarchy}), 
and our conclusions remain unchanged irrespective of the value of 
$\theta_{12}$ as long as (\ref{eqn:hierarchy}) holds. 

One can show \cite{mina} \cite{bbgk} that under the mass hierarchy 
(\ref{eqn:hierarchy}) the relation 
\begin{eqnarray}
U_{e3}^2\ll 1,
\label{eqn:e3}
\end{eqnarray}
must hold in order to have solar neutrino deficit under the constraints 
from the accelerator and the reactor experiments. In this setting it 
can also be demonstrated that the solar neutrino problem is indeed 
solved by a two-flavor framework in the MSW and the vacuum solutions
\cite{mina}.

In the present case the formulas in (\ref{eqn:3flavor}) become much 
simpler \cite{3nu}:
\begin{eqnarray}
&{ }&1-P(\nu_\alpha\rightarrow\nu_\alpha)=
4\sin^2\left({\Delta m_{21}^2L \over 4E}\right)U_{\alpha1}^2U_{\alpha2}^2
+4\sin^2\left({\Delta m_{31}^2L \over 4E}\right)
U_{\alpha3}^2(1-U_{\alpha3}^2)\nonumber\\
&{ }&P(\nu_\alpha\rightarrow\nu_\beta)=
-4\sin^2\left({\Delta m_{21}^2L \over 4E}\right)
U_{\alpha1}U_{\alpha2}U_{\beta1}U_{\beta2}
+4\sin^2\left({\Delta m_{31}^2L \over 4E}\right)
U_{\alpha3}^2U_{\beta3}^2.\nonumber\\
\label{eqn:prob1}
\end{eqnarray}
For the cases discussed below, we have $|\Delta m_{21}^2L/4E|\ll 1$
and the first term in each formula in (\ref{eqn:prob1}) can be
ignored.  Hence we get
\begin{eqnarray}
4U_{\alpha3}^2(1-U_{\alpha3}^2)&\le&\sin^22\theta_{\alpha}(\Delta m_{31}^2)
\quad{\rm (disappearance~experiments)}\nonumber\\
4U_{\alpha3}^2U_{\beta3}^2&\le&\sin^22\theta_{\alpha\beta}(\Delta m_{31}^2)
\quad{\rm (appearance~experiments)}
\label{eqn:prob2}
\end{eqnarray}
for negative results, respectively.

The data by the LSND group suggests that the allowed region for
the mass-squared difference $\Delta m_{31}^2$ is approximately
$\Delta m_{31}^2\gsim 5\times 10^{-2}$eV$^2$.  If we combine the
data on the same channel ${\overline \nu_\mu}\rightarrow
{\overline \nu_e}$ (and $\nu_\mu\rightarrow\nu_e$)
by the E776 group \cite{e776}, however, the region
$\Delta m_{31}^2\gsim 2.5$eV$^2$ seems to be almost excluded,
and we have
$5\times 10^{-2}$eV$^2 \lsim \Delta m_{31}^2 \lsim 2.5$eV$^2$
at 90\% confidence level.

As has been pointed out in Ref. \cite{3nu}, strong constraints
on the mixing angle come from the reactor experiment \cite{bugey}.
Using (\ref{eqn:prob2}) we have the constraint from the reactor
experiment \cite{bugey}
\begin{eqnarray}
\sin^22\theta_{13}=4U_{e3}^2(1-U_{e3}^2)\le
\sin^22\theta_{\rm Bugey}(\Delta m_{31}^2),
\label{eqn:bugey}
\end{eqnarray}
where $\sin^22\theta_{\rm Bugey}(\Delta m_{31}^2)$ stands for
the value of $\sin^22\theta$ on the boundary of the allowed
region in the ($\Delta m^2,\sin^22\theta$) plot in \cite{bugey}.
For the entire region 
$5\times 10^{-2}$eV$^2 \lsim \Delta m_{31}^2 \lsim 2.5$eV$^2$,
$\sin^22\theta_{\rm Bugey}(\Delta m_{31}^2)$ is small \cite{bugey}:
\begin{eqnarray}
{1 \over 50}\lsim\sin^22\theta_{\rm Bugey}(\Delta m_{31}^2)
\lsim{1 \over 10}.
\label{eqn:bugey2}
\end{eqnarray}
From (\ref{eqn:bugey}) we have either small $U_{e3}^2$ or
large $U_{e3}^2$, but the latter possibility is excluded
by (\ref{eqn:e3}).
Therefore we have
\begin{eqnarray}
s_{13}^2=U_{e3}^2\lsim{1 \over 4}\sin^22\theta_{\rm Bugey}
(\Delta m_{31}^2).
\end{eqnarray}

On the other hand, we have another constraint from the disappearance
experiment of $\nu_\mu$ \cite{cdhsw}
\begin{eqnarray}
4s_{23}^2c_{13}^2(1-s_{23}^2c_{13}^2)=4U_{\mu3}^2(1-U_{\mu3}^2)\le
\sin^22\theta_{\rm CDHSW}(\Delta m_{31}^2),
\label{eqn:cdhsw}
\end{eqnarray}
where $\sin^22\theta_{\rm CDHSW}(\Delta m_{31}^2)$ stands for
the value of $\sin^22\theta$ on the boundary of the allowed
region in the ($\Delta m^2,\sin^22\theta$) plot in \cite{cdhsw}.
The mixing angle in this case is constrained for
$0.7$eV$^2 \lsim \Delta m_{31}^2 \lsim 13$eV$^2$ as
\begin{eqnarray}
\sin^22\theta_{\rm CDHSW}(\Delta m_{31}^2) \lsim 0.2.
\label{eqn:cdhsw2}
\end{eqnarray}
If we consider the probability $P(\nu_\mu\rightarrow\nu_\mu)$
in the atmospheric neutrino experiments for the mass region above, 
we have small deviation of $P(\nu_\mu\rightarrow\nu_\mu)$ from unity 
\begin{eqnarray}
1-P(\nu_\mu\rightarrow\nu_\mu) \simeq
2 U_{\mu3}^2(1-U_{\mu3}^2)\lsim 0.1,
\label{eqn:mu3}
\end{eqnarray}
where we have averaged over rapid oscillations. 
(\ref{eqn:mu3}) is obviously inconsistent with the atmospheric 
neutrino observations \cite{kamioka1} \cite{kamioka2} \cite{imb} 
\cite{soudan2}, since we cannot have gross deficit of $\nu_\mu$ 
in this case.  In fact we have verified explicitly, by the same
calculation as in Ref. \cite{y}, that the region $\Delta m_{31}^2\gsim
0.47$eV$^2$ with the constraints (\ref{eqn:bugey}) and
(\ref{eqn:cdhsw}) is excluded at 95\% confidence level by the
Kamiokande multi-GeV data ($\chi^2\ge$17.6 for 3 degrees of freedom,
where $\chi_{\rm min}^2$=3.2, $(\chi^2-\chi_{\rm min}^2)/7\simeq 2.1$
implies 2$\sigma$).

So we are left with the region
$5\times 10^{-2}$eV$^2 \lsim \Delta m_{31}^2 \lsim 0.47$eV$^2$.
Applying the formula (\ref{eqn:prob1}) to the case of LSND,
we have 
\begin{eqnarray}
s_{23}^2\sin^22\theta_{13}=4U_{e3}^2 U_{\mu3}^2=
\sin^22\theta_{\rm LSND}(\Delta m_{31}^2),
\end{eqnarray}
where $\sin^22\theta_{\rm LSND}(\Delta m_{31}^2)$ stands for
the value of $\sin^22\theta$ within the allowed
region in the ($\Delta m^2,\sin^22\theta$) plot in \cite{lsnd},
and the LSND data indicates
\begin{eqnarray}
1.5\times 10^{-3}\lsim\sin^22\theta_{\rm LSND}(\Delta m_{31}^2)\le 1.
\end{eqnarray}
From (\ref{eqn:bugey}) and the constraint $s_{23}^2\le 1$, it follows that
\begin{eqnarray}
\sin^22\theta_{\rm LSND}(\Delta m_{31}^2)\le
\sin^22\theta_{\rm Bugey}(\Delta m_{31}^2),
\end{eqnarray}
so that we have
\begin{eqnarray}
\Delta m_{31}^2 \gsim 0.25{\rm eV}^2,
\end{eqnarray}
with
\begin{eqnarray}
1.5\times 10^{-3}\lsim\sin^22\theta_{\rm LSND}(\Delta_{31}^2)\lsim
3.8\times10^{-2},
\end{eqnarray}
where we have used the ($\Delta m^2,\sin^22\theta$) plots
in \cite{bugey} and \cite{lsnd}.
However, we have verified explicitly again
that the region $\Delta m^2\gsim 0.25$eV$^2$ is excluded
at 90\% confidence level by the Kamiokande multi-GeV data
($\chi^2\ge$15 for three degrees of
freedom, $(\chi^2-\chi_{\rm min}^2)/7\simeq 1.7$ implies 1.6$\sigma$).
Therefore, we conclude that the LSND data cannot be explained
by neutrino oscillations among three flavors, if all the
accelerator and reactor data as well as the Kamiokande
multi-GeV data are taken for granted.

The present LSND data allows conflicting interpretations either
as a possible evidence for neutrino oscillation \cite{lsnd},
or a stringent bound for the mixing parameters \cite{hill}.
It might be possible that the allowed region of the set 
of the parameters ($\Delta m^2$, $\sin^22\theta$) implied
by the LSND data changes in the future.
We have looked for the region of
($\Delta m^2$, $\sin^22\theta$) for $\nu_\mu\rightarrow\nu_e$ 
(or ${\overline \nu_\mu}\rightarrow{\overline \nu_e}$),
in which $\nu_\mu\rightarrow\nu_e$ oscillation is consistent with 
all the experiments (except LSND), including the Kamiokande multi-GeV 
data.  To obtain the ($\Delta m^2$, $\sin^22\theta_{\mu e}$) plot of the 
two-flavor analysis for general $\nu_\mu\rightarrow\nu_e$ 
oscillations with a mixing angle $\theta_{\mu e}(\Delta m_{31}^2)$,
we use the following correspondence between the rates for the 
$\nu_\mu\rightarrow e$ process in the three and the two-flavor 
frameworks:
\begin{eqnarray}
{N(\nu_\mu\rightarrow e;L)
\over N_{\mu e}(L)}&\simeq&
4U_{e3}^2 U_{\mu3}^2
\left\langle\sin^2\left({\Delta m_{31}^2L \over 4E}\right)
\right\rangle_{\mu\rightarrow e}\nonumber\\
&=&s_{23}^2\sin^22\theta_{13}^2
\left\langle\sin^2\left({\Delta m_{31}^2L \over 4E}\right)
\right\rangle_{\mu\rightarrow e}\nonumber\\
&\equiv&\sin^22\theta_{\mu e}(\Delta m_{31}^2)
\left\langle\sin^2\left({\Delta m_{31}^2L \over 4E}\right)
\right\rangle_{\mu\rightarrow e},
\label{eqn:correspondence}
\end{eqnarray}
where we have used the hierarchical condition (\ref{eqn:hierarchy}).
From (\ref{eqn:bugey}) and (\ref{eqn:correspondence}) we have 
\begin{eqnarray}
\sin^22\theta_{\mu e}(\Delta m_{31}^2)
&\simeq&s_{23}^2 \sin^22\theta_{13}\nonumber\\
&\le&
s_{23}^2 \sin^22\theta_{\rm Bugey}(\Delta m_{31}^2).
\label{eqn:emu}
\end{eqnarray}
The Kamiokande multi-GeV data has better fit for larger values
of $\sin^22\theta_{23}$, so (\ref{eqn:emu}) shows that
$\sin^22\theta_{\mu e}(\Delta m_{31}^2)$
cannot be larger than $\sin^22\theta_{\rm Bugey}(\Delta m_{31}^2)/2$,
if $\sin^22\theta_{13}\ll 1$\footnote{This is not the case for
$\Delta m_{31}^2 \lsim 3\times 10^{-2}$eV$^2$, where
$1-P(\nu_\mu\rightarrow\nu_\mu)=4s_{23}^2c_{13}^2(1-s_{23}^2c_{13}^2)
\sim {\cal O} (1)$ becomes possible even for $\theta_{23}<\pi/4$.}.
The result is shown in Fig.1.

\vglue 0.5truecm
(Insert Fig.1 here.)
\vglue 0.5truecm

The region suggested by the LSND
experiment \cite{lsnd} is close to the 90\%CL region
obtained in our analysis.  If the 20\% systematic uncertainty
mentioned in Ref. \cite{lsnd} shifts
the allowed region of LSND in the direction of smaller mixing,
there may be a chance that all the neutrino
anomalies are explained by three flavor neutrino oscillations.
Hopefully further data from the LSND group will clarify
the situation.

\vskip 0.2in
\noindent
{\Large{\bf Acknowledgement}}
\vskip 0.1in

One of the authors (O.Y.) would like to thank K.S. Babu, P.I. Krastev,
C.N. Leung, and A. Smirnov for discussions.
We would like to thank K. Iida for collaboration in the early stages 
of this work and members of the Physics Department of Yale University 
for their hospitality during our stay while a part of this work was done.
This research was supported in part by a Grant-in-Aid for Scientific
Research, Ministry of Education, Science and Culture, 
\#05302016, \#05640355, and \#07044092.

\newpage
\noindent
{\Large{\bf Figures}}

\begin{description}
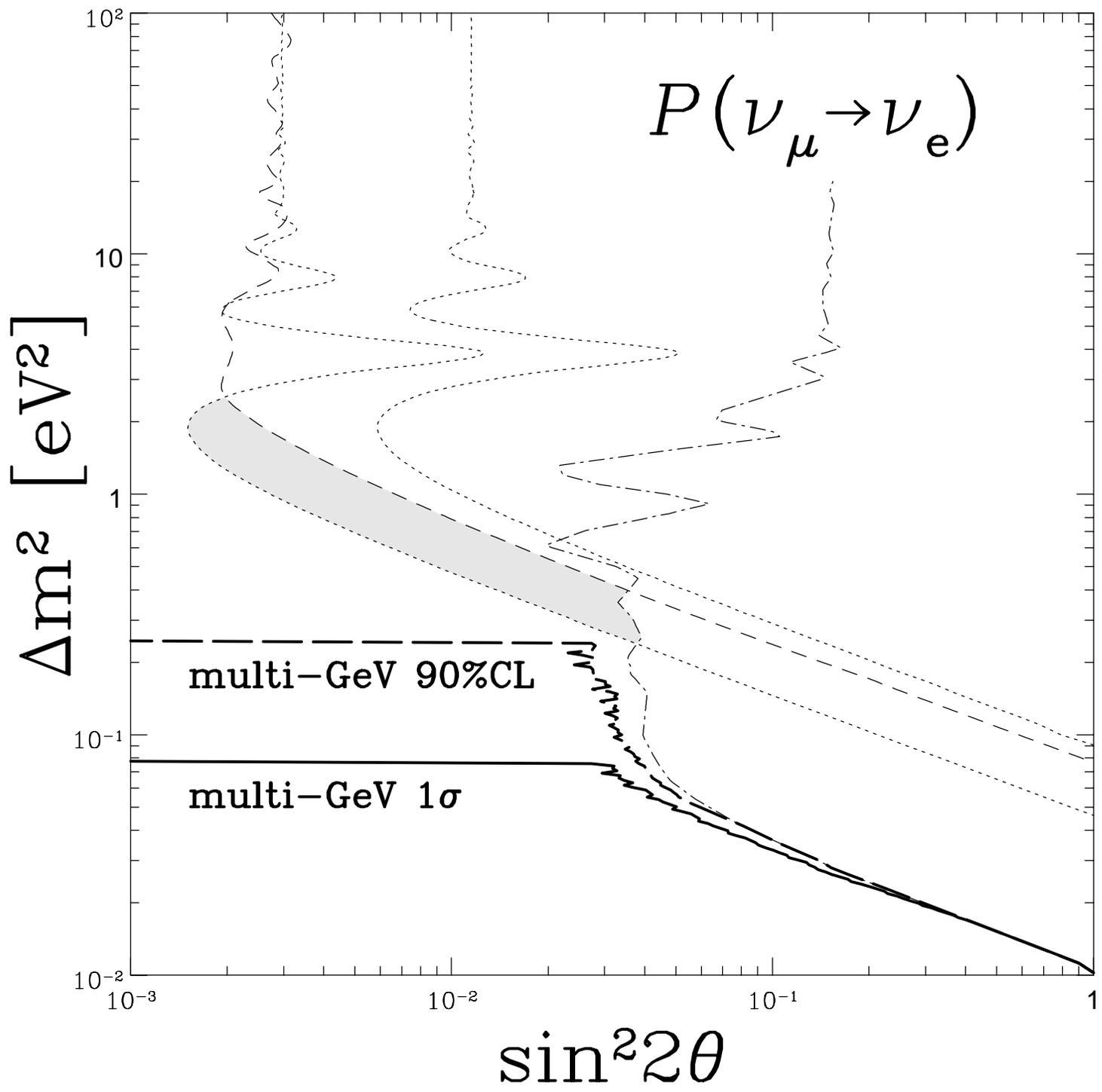
\item[Fig.1] Below the solid and dashed lines
are the regions for $\nu_\mu\rightarrow\nu_e$ (or
${\overline \nu_\mu}\rightarrow{\overline \nu_e}$)
oscillation which are compatible with all the experiments
(except LSND), including the atmospheric
multi-GeV data of Kamiokande at 68\%CL (solid) and 90\%CL (dashed),
respectively.  To get this plot of the two-flavor mixing
parameters, we have used the correspondence
$\sin^22\theta\equiv\sin^22\theta_{\mu e}(\Delta m_{31}^2)=
s_{23}^2\sin^22\theta_{13}^2$ (cf. (\ref{eqn:correspondence})).
The shadowed area stands for the region allowed by all the
accelerator and reactor experiments including LSND
(The dotted, dashed and dot-dashed lines stand for the
LSND \cite{lsnd}, E776 \cite{e776},
and Bugey \cite{bugey} experiments, respectively).  
\end{description}

\end{document}